\documentclass[fleqn,twoside]{article}
\usepackage{espcrc2}
\usepackage[figuresright]{rotating}

\usepackage{graphicx}
\newcommand{\AmS}{{\protect\the\textfont2
  A\kern-.1667em\lower.5ex\hbox{M}\kern-.125emS}}

\title{Study of Quark Confinement in Baryons with Lattice QCD}

\author{
Hideo Suganuma\address[titech]{
\vspace{-0.1cm}
Faculty of Science, Tokyo Institute of Technology,
Ohokayama 2-12-1, Meguro, Tokyo 152-8551, Japan}, 
Toru T. Takahashi\address[yitp]{
\vspace{-0.1cm}
Yukawa Institute for Theoretical Physics, Kyoto University, 
Kitashirakawa, Sakyo 606-8502, Japan},
Fumiko Okiharu\address[nihon]{
\vspace{-0.1cm}
Department of Physics, Nihon University, 1-8 Kanda-Surugadai, Chiyoda, Tokyo 101, Japan}
and 
Hiroko Ichie\addressmark[titech]
}

\begin{document}

\begin{abstract}
In SU(3) lattice QCD, 
we perform the detailed study for the ground-state three-quark (3Q) potential $V_{\rm 3Q}^{\rm g.s.}$ 
and the 1st excited-state 3Q potential $V_{\rm 3Q}^{\rm e.s.}$, i.e., 
the energies of the ground state and the 1st excited state 
of the gluon field in the presence of the static three quarks.
From the accurate calculation for 
more than 300 different patterns of 3Q systems, the static ground-state 3Q potential 
$V_{\rm 3Q}^{\rm g.s.}$ is found to be well described 
by the Coulomb plus Y-type linear potential (Y-Ansatz) within 1\%-level deviation.
As a clear evidence for Y-Ansatz, 
Y-type flux-tube formation is actually observed on the lattice in maximally-Abelian projected QCD.
For about 100 patterns of 3Q systems, 
we calculate the 1st excited-state 3Q potential $V_{\rm 3Q}^{\rm e.s.}$, and find 
a large gluonic-excitation energy $\Delta E_{\rm 3Q} \equiv V_{\rm 3Q}^{\rm e.s.}-V_{\rm 3Q}^{\rm g.s.}$ 
of about 1 GeV, which gives a physical reason of 
the success of the quark model even without gluonic excitations.
We present also the first study for the penta-quark potential $V_{\rm 5Q}$ 
in lattice QCD, and find that 
$V_{\rm 5Q}$ is well described by the sum of the OGE Coulomb plus multi-Y type linear potential.
\vspace{-0.5cm}
\end{abstract}

\maketitle

\section{Introduction}

Quantum chromodynamics (QCD), the SU(3) gauge theory,  
was first proposed by Yoichiro Nambu \cite{N66} in 1966 as a candidate for 
the fundamental theory of the strong interaction, 
just after the introduction of the ``new" quantum number, ``color" \cite{HN65}. 
In spite of its simple form, QCD creates thousands of hadrons and leads to various interesting nonperturbative phenomena 
such as color confinement \cite{conf2003} and dynamical chiral-symmetry breaking \cite{NJL61}.
Even now, it is very difficult to deal with QCD due to its strong-coupling nature in the infrared region.

In recent years, 
the lattice QCD Monte Carlo calculation becomes a reliable and useful method
for the analysis of nonperturbative QCD \cite{R97},  
which indicates an important direction in the hadron physics.
In this paper, using lattice QCD, we study the inter-quark potential in detail \cite{TMNS01,TSNM02,TS03,TMNS99}.

In general, the three-body force is regarded as a residual interaction in most fields in physics.
In QCD, however, the three-body force among three quarks is 
a ``primary" force reflecting the SU(3) gauge symmetry.
In fact, the three-quark (3Q) potential is directly responsible 
for the structure and properties of baryons, 
similar to the relevant role of the Q-$\bar{\rm Q}$ potential for meson properties, 
and both the Q-$\bar{\rm Q}$ potential and 
the 3Q potential are equally important fundamental quantities in QCD.
Furthermore, the 3Q potential is the key quantity to clarify the quark confinement in baryons.
However, in contrast to the Q-$\bar{\rm Q}$ potential \cite{R97},
there was almost no lattice QCD study for the 3Q potential before our study in 1999 \cite{TMNS99},
in spite of its importance in the hadron physics. 

\section{The Ground-State 3Q Potential in QCD}

The Q-$\bar {\rm Q}$ potential is known to be well described with 
the inter-quark distance $r$ as \cite{R97,TMNS01,TSNM02} 
\begin{eqnarray}
V_{\rm Q \bar Q}(r)=-\frac{A_{\rm Q\bar Q}}{r}+\sigma_{\rm Q \bar Q}r+C_{\rm Q\bar Q}.
\label{eqn:QQpotential}
\end{eqnarray}
As for the 3Q potential form, we note two theoretical arguments 
at short and long distance limits.

\vspace{0.2cm}

\noindent
1. At the short distance, perturbative QCD is applicable,
and therefore 3Q potential is expressed as the sum of the two-body Coulomb potential 
originating from the one-gluon-exchange process.

\vspace{0.2cm}

\noindent
2. At the long distance, the strong-coupling expansion of QCD is plausible, and it 
leads to the flux-tube picture \cite{KS75CKP83}.
For the 3Q system, there appears a junction which connects the three flux-tubes from the three quarks, 
and Y-type flux-tube formation is expected \cite{TMNS01,TSNM02,KS75CKP83}.

\vspace{0.2cm}

Then, we theoretically conjecture the functional form of the 3Q potential 
as the Coulomb plus Y-type linear potential, i.e., Y-Ansatz,
\begin{eqnarray}
V_{\rm 3Q}^{\rm g.s.}=-A_{\rm 3Q}\sum_{i<j}\frac1{|r_i-r_j|}+
\sigma_{\rm 3Q}L_{\rm min}+C_{\rm 3Q},
\label{eqn:3Qpotential}
\end{eqnarray}
where $L_{\rm min}$ is the minimal value of the total flux-tube length.
Of course, it is nontrivial that these simple arguments on UV and IR limits of QCD hold for the intermediate region. 
Then, we study the 3Q potential in lattice QCD.
Note that the lattice QCD calculation is completely independent of any Ansatz for the potential form.

\subsection{The Three-Quark Wilson Loop}

Similar to the Q-$\bar {\rm Q}$ potential calculated with the Wilson loop, 
the 3Q potential can be calculated  with 
the 3Q Wilson loop \cite{TMNS01,TSNM02,TS03} 
defined as 
\begin{eqnarray}
W_{\rm 3Q}\equiv \frac1{3!}\epsilon_{abc}\epsilon_{a'b'c'}U_1^{aa'}U_2^{bb'}U_3^{cc'}
\end{eqnarray}
with
$U_k\equiv P\exp\{ig\int_{\Gamma_k}dx_\mu A^\mu(x)\}$ in Fig.1.
The 3Q Wilson loop physically means that
a color-singlet gauge-invariant 3Q state is created at $t=0$ and 
is annihilated at $t=T$ with the three quarks spatially fixed for $0<t<T$.

\begin{figure}[hb]
\vspace{-0.65cm}
\centering
\includegraphics[height=3.4cm]{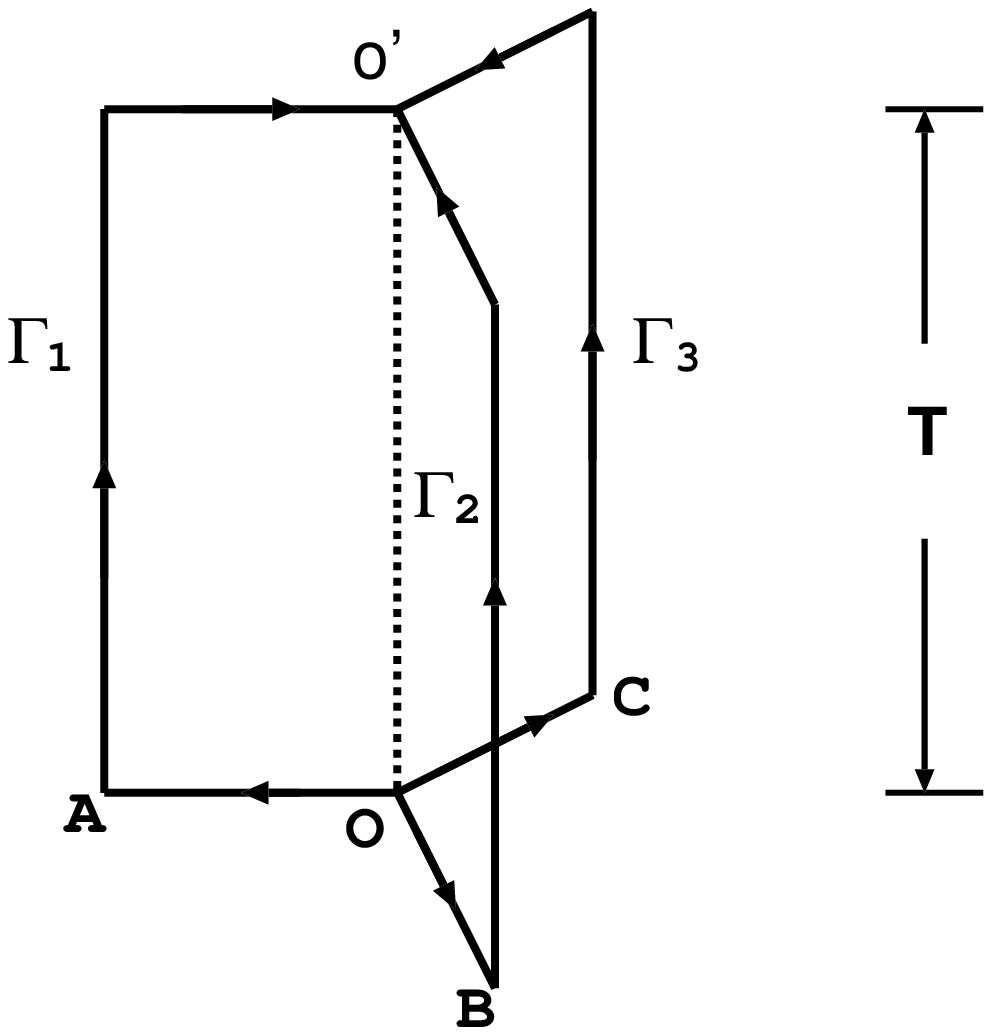}
\vspace{-1.05cm}
\caption{The 3Q Wilson loop $W_{\rm 3Q}$.
}
\label{fig1}
\vspace{-0.7cm}
\end{figure}

The vacuum expectation value of the 3Q Wilson loop is expressed as 
\begin{eqnarray}
\langle W_{\rm 3Q} \rangle =\sum_{n=0}^{\infty} C_n \exp(-V_n T),
\end{eqnarray}
where $V_n$ denotes the $n$th energy of the gauge-field configuration 
in the presence of the spatially-fixed three quarks \cite{TMNS01,TSNM02,TS03}.

While $V_n$ depends only on the 3Q location,
$C_n$ depends on the operator choice for  the 3Q state.
For the accurate calculation, we need to 
reduce the excited-state component, i.e., $C_n(n\ge 1)$, in the 3Q state prepared at $t=0$ and $T$.

\subsection{The Smearing Method}

The smearing is a useful method to construct the quasi-ground-state operator in lattice QCD 
in a gauge-invariant manner, 
and is actually successful for the ground-state Q-$\bar {\rm Q}$ potential \cite{TSNM02}.
(Note that the smearing is just a method to 
construct the operator, and hence 
it never changes the gauge configuration  
unlike the cooling.)

The smeared link-variable $\bar U_\mu(s)$ includes a spatial extension, 
and the smeared ``line'' expressed with $\bar U_\mu(s)$ 
corresponds to a Gaussian-distributed ``flux-tube'' 
in terms of the original link-variable \cite{TSNM02}.
Therefore, the properly smeared line 
is expected to resemble the ground-state flux-tube. 

Then, through the selection of the properly smeared 3Q Wilson loop $\langle W_{\rm 3Q}[\bar U_\mu(s)]\rangle$, 
we can construct the ground-state-dominant 3Q operator  
for the accurate measurement of 
the ground-state 3Q potential \cite{TMNS01,TSNM02,TS03}.

\subsection{Lattice QCD results for 3Q Potential}

For more than 300 different patterns of spatially-fixed 3Q systems, 
we perform the thorough calculation  
of the ground-state potential $V_{\rm 3Q}^{\rm g.s.}$ 
in SU(3) lattice QCD with the standard plaquette action 
with $12^3\times 24$ at $\beta=5.7$ and 
with $16^3\times 32$ at $\beta$=5.8 and 6.0 at the quenched level.
For the accurate measurement, we 
use the smearing method and construct the ground-state-dominant 
3Q operator \cite{TMNS01,TSNM02,TS03}.

To conclude, we find that the static ground-state 3Q potential $V_{\rm 3Q}^{\rm g.s.}$
is well described by the Coulomb plus Y-type linear potential (Y-Ansatz)  
within 1\%-level deviation \cite{TMNS01,TSNM02}.

\subsection{Other Studies on the 3Q Potential}

To clarify the current status of the 3Q potential, 
we introduce recent works of other groups.

de Forcrand's group, who once supported $\Delta$-Ansatz in lattice QCD \cite{AdFT02}, 
seems to change their opinion from $\Delta$-Ansatz to Y-Ansatz \cite{JdF04}. 

Kuzmenko and Simonov also showed that the Delta-shape is impossible 
from gauge-invariance point of view, 
and the Y-shaped configuration is the only possible for the three-quark 
system \cite{KS03}.

One of the theoretical basis of $\Delta$-Ansatz was Cornwall's conjecture 
based on the vortex vacuum model \cite{C96}.
Recently, Cornwall re-examined his previous work and found that 
the correct answer is Y-Ansatz instead of $\Delta$-Ansatz \cite{C04}.

In this way, Y-Ansatz for the static 3Q potential
seems almost settled both in lattice QCD and in analytic framework. 

\subsection{Y-type Flux-Tube Formation}

Recently, as a clear evidence for Y-Ansatz, 
Y-type flux-tube formation is actually observed 
in maximally-Abelian (MA) projected lattice QCD 
from the measurement of the action density  
in the spatially-fixed 3Q system \cite{IBSS03,STI04}. 
(See Fig.2.)

\begin{figure}[hb]
\vspace{-0.8cm}
\hspace{-0.25cm}\includegraphics[width=3.2in]{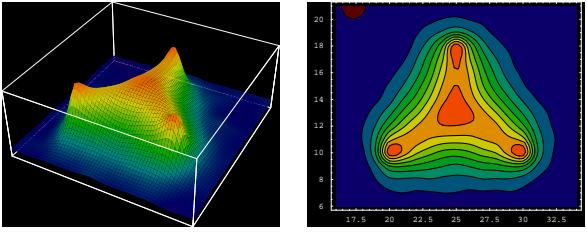}
\vspace{-1.2cm}
\caption{
The lattice QCD result for Y-type flux-tube formation 
in the spatially-fixed 3Q system in MA projected QCD.
The distance between the junction and each quark is about 0.5 fm.
}
\label{fig2}
\vspace{-1cm}
\end{figure}

\section{Penta-Quark Potential in Lattice QCD}

Motivated by the recent discovery of the penta-quark baryon $\Theta^+(1540)$, 
we perform the first study of the static penta-quark (5Q) potential $V_{\rm 5Q}$ 
in SU(3) lattice QCD with $\beta$=6.0 and $16^3 \times$ 32 
at the quenched level. We investigate the QQ-$\bar {\rm Q}$-QQ configuration 
as shown in Fig.3.
With the smearing method \cite{TMNS01,TSNM02,TS03} 
to enhance the ground-state component, 
we accurately calculate the 5Q potential $V_{\rm 5Q}$
from the 5Q Wilson loop $\langle W_{\rm 5Q} \rangle$ as shown in Fig.4 
in a gauge-invariant manner.

\begin{figure}[h]
\centerline{
\includegraphics[height=2.2cm]{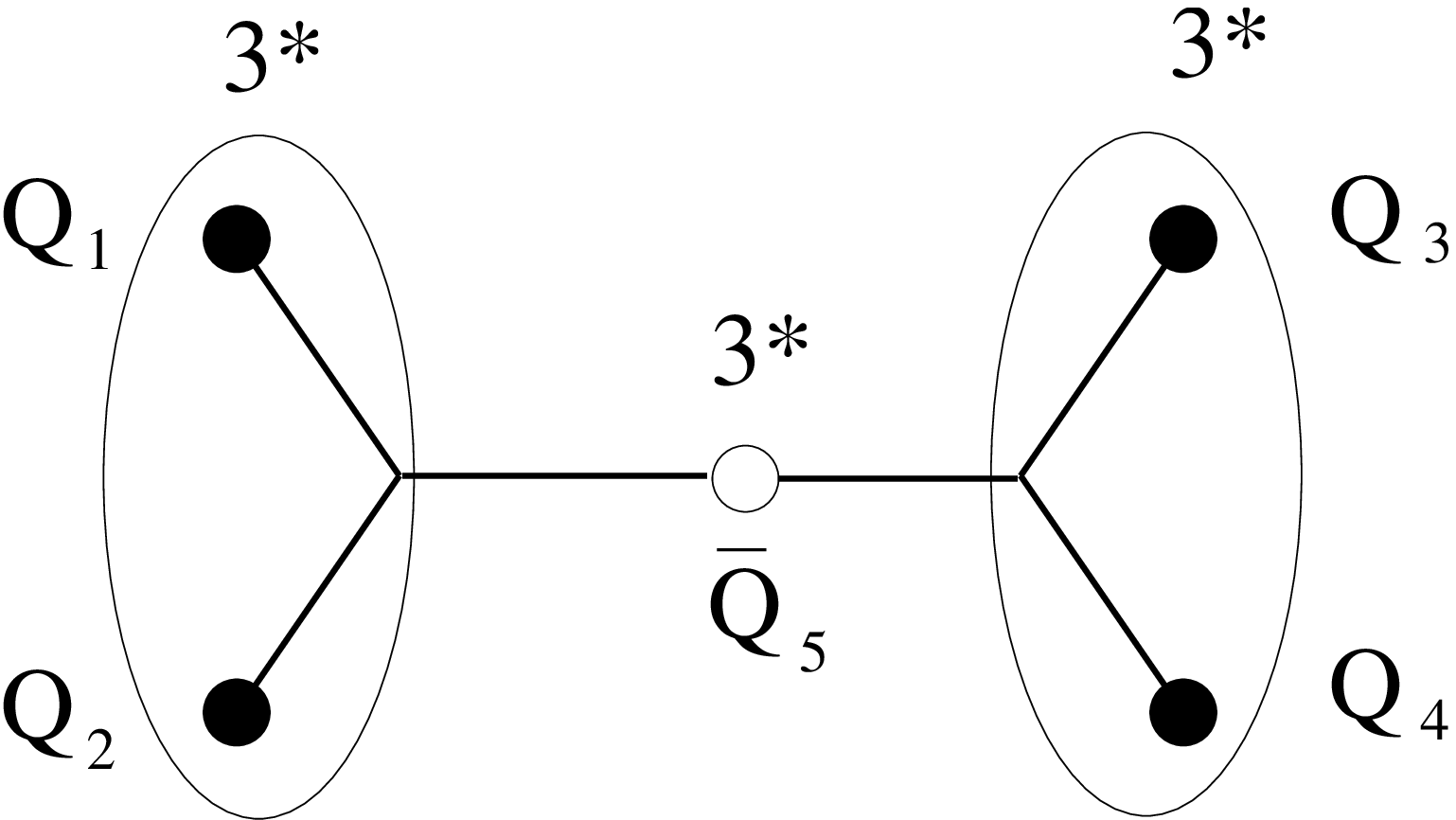}
\hspace{0.2cm}
\includegraphics[height=2.2cm]{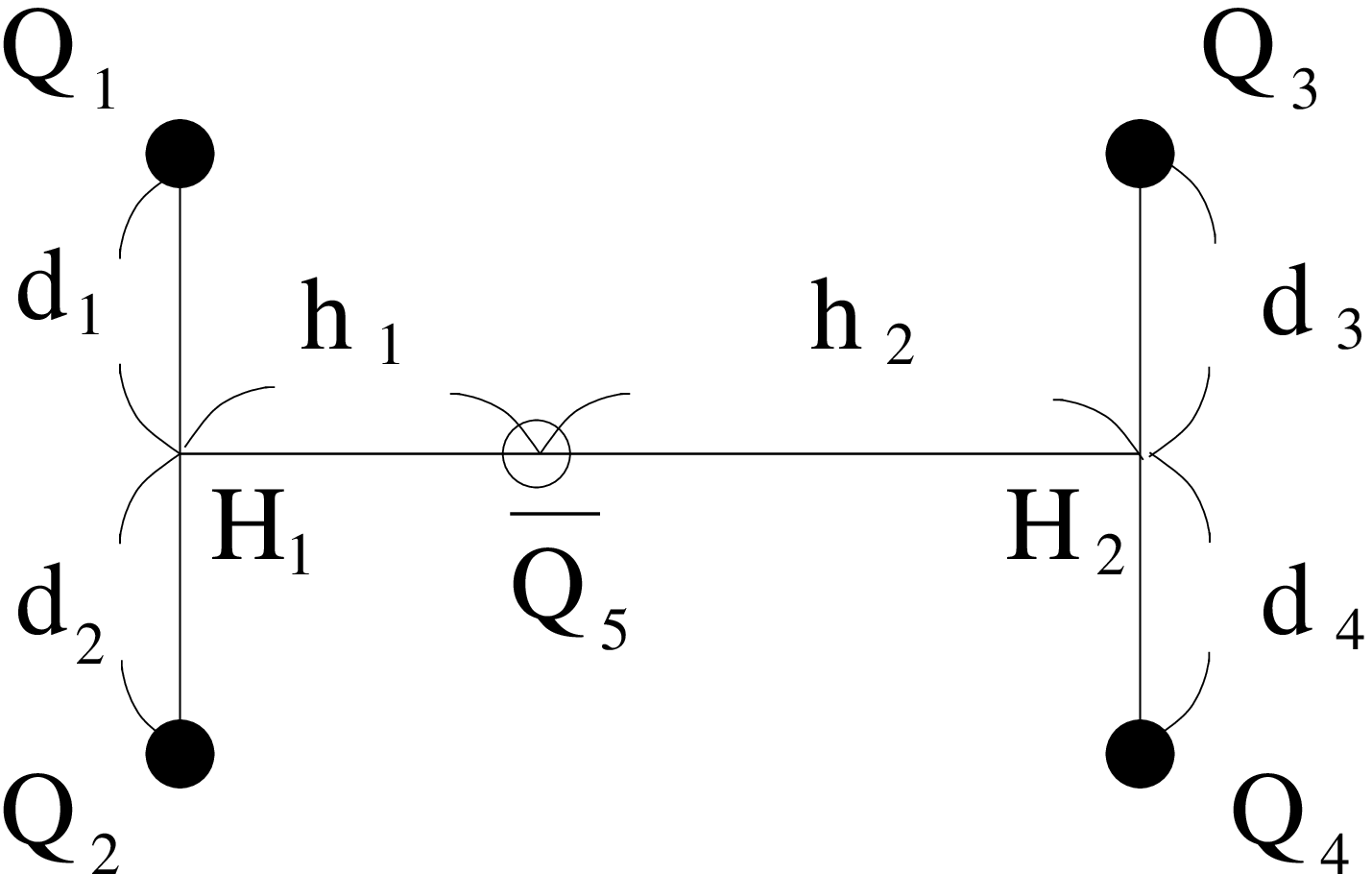}
}
\vspace{-0.7cm}
\caption{
(a) The QQ-$\bar {\rm Q}$-QQ type configuration for the penta-quark system.
The two QQ clusters belong to the {\bf 3}* representation of the color SU(3).
(b) A planar configuration of 
the penta-quark system.
We take $d_1=d_2=d_3=d_4\equiv d$.
}
\label{fig3}
\vspace{-0.7cm}
\end{figure}

\begin{figure}[h]
\hspace{0.4cm} \includegraphics[width=2.3in]{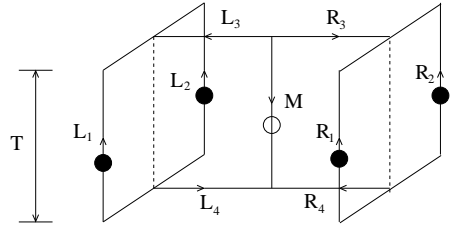}
\vspace{-0.8cm}
\caption{
The penta-quark (5Q) Wilson loop $W_{\rm 5Q}$ for the 5Q potential $V_{\rm 5Q}$.
The contours $M, L_i, R_i (i=3,4)$ are line-like and $L_i, R_i (i=1,2)$ are staple-like.
The 5Q gauge-invariant state is generated at $t=0$ and is annihilated at $t=T$. 
}
\label{fig4}
\vspace{-0.5cm}
\end{figure}

\begin{figure}[h]
\begin{center}
\rotatebox{-90}{\includegraphics[height=2.5in]{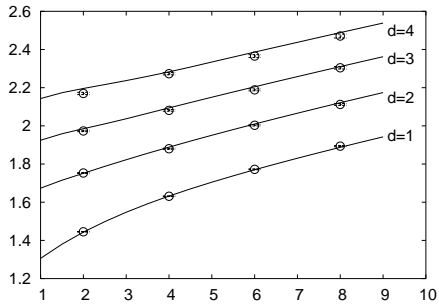}}
\end{center}
\vspace{-1.4cm}
\caption{
Lattice QCD results of the penta-quark potential $V_{\rm 5Q}$ 
for the planar 5Q configuration with $h_1=h_2\equiv h$ in Fig.3 in the lattice unit. 
Each 5Q system is labeled by $d$ and $h$. 
The symbols denote the lattice data, and the curves 
the theoretical form of the OGE plus multi-Y Ansatz. 
}
\label{fig5}
\vspace{-0.5cm}
\end{figure}

We find that the 5Q potential $V_{\rm 5Q}$ is well described by 
the sum of one-gluon-exchange (OGE) Coulomb term and multi-Y type linear term, which we call 
the ``OGE plus multi-Y Ansatz", 
\begin{eqnarray}
\hspace{-0.7cm}
&&V_{\rm 5Q}
=\frac{g^2}{4\pi} \sum_{i<j} \frac{T^a_i T^a_j}{|{\bf r}_i-{\bf r}_j|}+\sigma_{\rm 5Q} L_{\rm min}+C_{\rm 5Q} 
\label{pentaform}
\\
\hspace{-0.7cm}
&&=-A_{\rm 5Q}\{ ( \frac1{r_{12}}  + \frac1{r_{34}}) 
+\frac12(\frac1{r_{15}} +\frac1{r_{25}} +\frac1{r_{35}} +\frac1{r_{45}}) \nonumber \\
\hspace{-0.7cm}
&&+\frac14(\frac1{r_{13}} +\frac1{r_{14}} +\frac1{r_{23}} +\frac1{r_{24}}) \}
+\sigma_{\rm 5Q} L_{\rm min}+C_{\rm 5Q} \nonumber
\end{eqnarray}
with $r_{ij}\equiv|{\bf r}_i-{\bf r}_j|$ and 
$L_{\rm min}$ being the minimal length of the flux-tube linking five quarks.
Note that there appear three kinds of Coulomb coefficients ($A_{\rm 5Q}$, $\frac12 A_{\rm 5Q}$, $\frac14 A_{\rm 5Q}$) in the penta-quark system, 
while only one Coulomb coefficient, $A_{\rm Q\bar {\rm Q}}$ or $A_{\rm 3Q}$, appears in the Q$\bar{\rm Q}$ or the 3Q system. 
(In Eq.(\ref{pentaform}), $A_{\rm 5Q}$ corresponds to $A_{\rm 3Q}$ or $\frac12 A_{\rm Q\bar Q}$ in terms of the OGE result.) 

Figure 5 shows the lattice QCD results for the 5Q potential $V_{\rm 5Q}$.
The symbols denote the lattice data, and the curves denote the theoretical form 
of the OGE plus multi-Y Ansatz 
with ($A_{\rm 5Q}$,$\sigma_{\rm 5Q}$) fixed to be ($A_{\rm 3Q}$,$\sigma_{\rm 3Q}$) in 
the 3Q potential $V_{\rm 3Q}$ in Ref.\cite{TSNM02}. 
(Note that there is no adjustable parameter for the theoretical curves besides an irrelevant constant $C_{\rm 5Q}$, 
since $A_{\rm 5Q}$ and $\sigma_{\rm 5Q}$ are fixed.)
One finds a good agreement between the lattice data and the OGE plus multi-Y Ansatz.

We note that the multi-quark system including four or more quarks can take a three-dimensional shape, 
while the Q$\bar {\rm Q}$ and the 3Q systems can take only planar configuration. 
Then, we investigate also the twisted 5Q configuration, and find that 
the planar and the twisted 5Q configurations are almost degenerate.
Then, no special configuration is favored in the 5Q system in terms of the energy,  
and general 5Q systems tend to take a three-dimensional configuration.

From the comparison with the Q$\bar {\rm Q}$ 
and the 3Q potentials \cite{TMNS01,TSNM02,TS03}, the universality of the string tension 
and the OGE result are found among Q$\bar {\rm Q}$, 3Q and 5Q systems as 
\begin{eqnarray}
\sigma_{\rm Q\bar{\rm Q}}\simeq \sigma_{\rm 3Q} \simeq \sigma_{\rm 5Q}, \quad
\frac12A_{\rm Q\bar{\rm Q}}\simeq A_{\rm 3Q} \simeq A_{\rm 5Q}.
\end{eqnarray} 
This result supports the flux-tube picture 
on the confinement mechanism even for the multi-quark system.

\section{Gluonic Excitations in the 3Q System}

In 1969, Y.~Nambu first pointed out the string picture for hadrons \cite{N6970}. 
Since then, the string picture has been one of the most important pictures for hadrons 
and has provided many interesting ideas in the wide region of the particle physics.

For instance, the hadronic string creates infinite number of hadron resonances as the vibrational modes, 
and these excitations lead to the Hagedorn ``ultimate" temperature \cite{H65},
which gives an interesting  theoretical picture for 
the QCD phase transition.

For the real hadrons, of course, the hadronic string is to have a spatial extension like the flux-tube, 
as the result of one-dimensional squeezing of the color-electric flux 
in accordance with color confinement \cite{N74}. 
Therefore, the vibrational modes of the hadronic flux-tube should be much more complicated, and 
the analysis of the excitation modes is important to clarify the underlying picture for real hadrons. 

In the language of QCD, such non-quark-origin excitation is called as the ``gluonic excitation", and 
is physically interpreted as the excitation of 
the gluon-field configuration in the presence of the quark-antiquark pair or the three quarks. 

In the hadron physics, the gluonic excitation is one of the interesting phenomena 
beyond the quark model, and relates to the hybrid hadrons such as $q\bar qG$ and $qqqG$. 
In particular, the hybrid meson includes the exotic hadrons with 
$J^{PC}=0^{--},0^{+-},1^{-+},2^{+-},\cdots$,
which cannot be constructed within the simple quark model.

In this section, we study the excited-state 3Q potential and the gluonic excitation 
using lattice QCD \cite{TS03}, 
to get deeper insight on these excitations beyond the hypothetical models
such as the string and the flux-tube models.
Here, the excited-state 3Q potential is 
the energy of the excited state of the gluon-field configuration 
in the presence of the static three quarks, and 
the gluonic-excitation energy is expressed as  
the energy difference between the ground-state 3Q potential  
and the excited-state 3Q potential.

\subsection{General Formalism}

We present the formalism to extract the excited-state potential \cite{TS03}. 
For the simple notation, the ground state is regarded as the ``0th excited state''.
For the physical eigenstates of the QCD Hamiltonian $\hat H$ 
for the spatially-fixed 3Q system, we denote 
the $n$th excited state by $|n \rangle$ ($n=0,1,2,\cdots$).
Since the three quarks are spatially fixed in this case, 
the eigenvalue of $\hat H$ is expressed by a static potential
as $\hat H|n\rangle=V_n|n\rangle$, where $V_n$ denotes the $n$th excited-state 3Q potential. 
Note that both $V_n$ and $|n \rangle $ are universal 
physical quantities relating to the QCD Hamiltonian $\hat H$.
In fact, $V_n$ depends only on the 3Q location, and 
$|n \rangle $ satisfies the orthogonal condition as $\langle m|n \rangle=\delta_{mn}$.

Suppose that $|\Phi_k \rangle \ (k=0,1,2,\cdots)$ are arbitrary given 
independent spatially-fixed 3Q states. 
In general, each 3Q state $|\Phi_k \rangle$ can be expressed by 
a linear combination of 3Q physical eigenstates 
$|n\rangle $,
\begin{eqnarray}
|\Phi_k \rangle =c_0^k|0\rangle+c_1^k|1\rangle+c_2^k|2\rangle+\cdots.
\end{eqnarray}
Here, the coefficients $c_n^k$ depend on the selection of 
$|\Phi_k \rangle$, and hence they are not universal quantities.

The Euclidean-time evolution of the 3Q state $|\Phi_k(t)\rangle$ is 
expressed with the operator $e^{-\hat Ht}$, which corresponds to 
the transfer matrix in lattice QCD. 
The overlap $\langle \Phi_j(T)|\Phi_k(0)\rangle$ is given by 
the 3Q Wilson loop with the initial state $|\Phi_k\rangle$ 
at $t=0$ and the final state $|\Phi_j\rangle$ at $t=T$, 
and is expressed in the Euclidean Heisenberg picture as 
\begin{eqnarray}
\hspace{-0.7cm} &&W^{jk}_T \equiv 
\langle \Phi_j|W_{\rm 3Q}(T)|\Phi_k\rangle
=\langle\Phi_j|e^{-\hat HT}|\Phi_k\rangle \nonumber\\
\hspace{-0.7cm} &&=\sum_{m=0}^\infty \sum_{n=0}^\infty \bar c_m^j c_n^k
\langle m|e^{-\hat HT}|n \rangle 
=\sum_{n=0}^\infty \bar c_n^j c_n^k e^{-V_nT}.~~~~
\end{eqnarray}
Using the matrix $C$ satisfying $C^{nk} =c_n^k$
and the diagonal matrix $\Lambda_T$ as $\Lambda_T^{mn}=e^{-V_nT}\delta^{mn}$, 
we rewrite the above relation as 
\begin{equation}
W_T=C^\dagger \Lambda_T C.
\label{WandLambda}
\end{equation}
Note here that $C$ is not a unitary matrix, and hence this relation 
does not mean the simple diagonalization by the unitary transformation.

Since we are interested in the 3Q potential $V_{n}$ 
in $\Lambda_T$ rather than the non-universal matrix $C$, 
we single out $V_{n}$ from the 3Q Wilson loop $W_T$ as
\begin{eqnarray}
W^{-1}_TW_{T+1}=C^{-1}{\rm diag}(e^{-V_0},e^{-V_1},\cdots)C,
\end{eqnarray}
which is a similarity transformation.
Then, $e^{-V_n}$ can be obtained as the eigenvalues of the matrix 
$W_T^{-1}W_{T+1}$, i.e., solutions of the secular equation, 
\begin{eqnarray}
{\rm det}\{W_T^{-1}W_{T+1}-t{\bf 1}\}=\prod_{n}(e^{-V_n}-t)=0.
\label{secular}
\end{eqnarray}
Thus, the 3Q potential $V_n$ can be obtained from the matrix  $W_T^{-1}W_{T+1}$.

In the practical calculation, we prepare $N$ independent sample states 
$|\Phi_k \rangle \ (k=0,1,\cdots,N-1)$. 
By choosing appropriate states $|\Phi_k \rangle$ 
so as not to include highly excited-state components, 
the physical states $|n \rangle$ can be truncated as $0\le n \le N-1$.
Then, $W_T$, $C$ and $\Lambda_T$ are reduced into $N\times N$ matrices, 
and the secular equation (\ref{secular}) becomes the $N$th 
order equation. 

\subsection{Lattice QCD for Gluonic Excitations}

For about 100 different patterns of spatially-fixed 3Q systems, 
we calculate the excited-state potential 
using SU(3) lattice QCD with $16^3\times 32$ at $\beta$=5.8 and 6.0 at the quenched level \cite{TS03}. 
In Fig.6, we show the 1st excited-state 3Q potential $V_{\rm 3Q}^{\rm e.s.}$ and 
the ground-state potential $V_{\rm 3Q}^{\rm g.s.}$.

\begin{figure}[hb]
\vspace{-0.6cm}
\centerline{
\includegraphics[height=3.3cm]{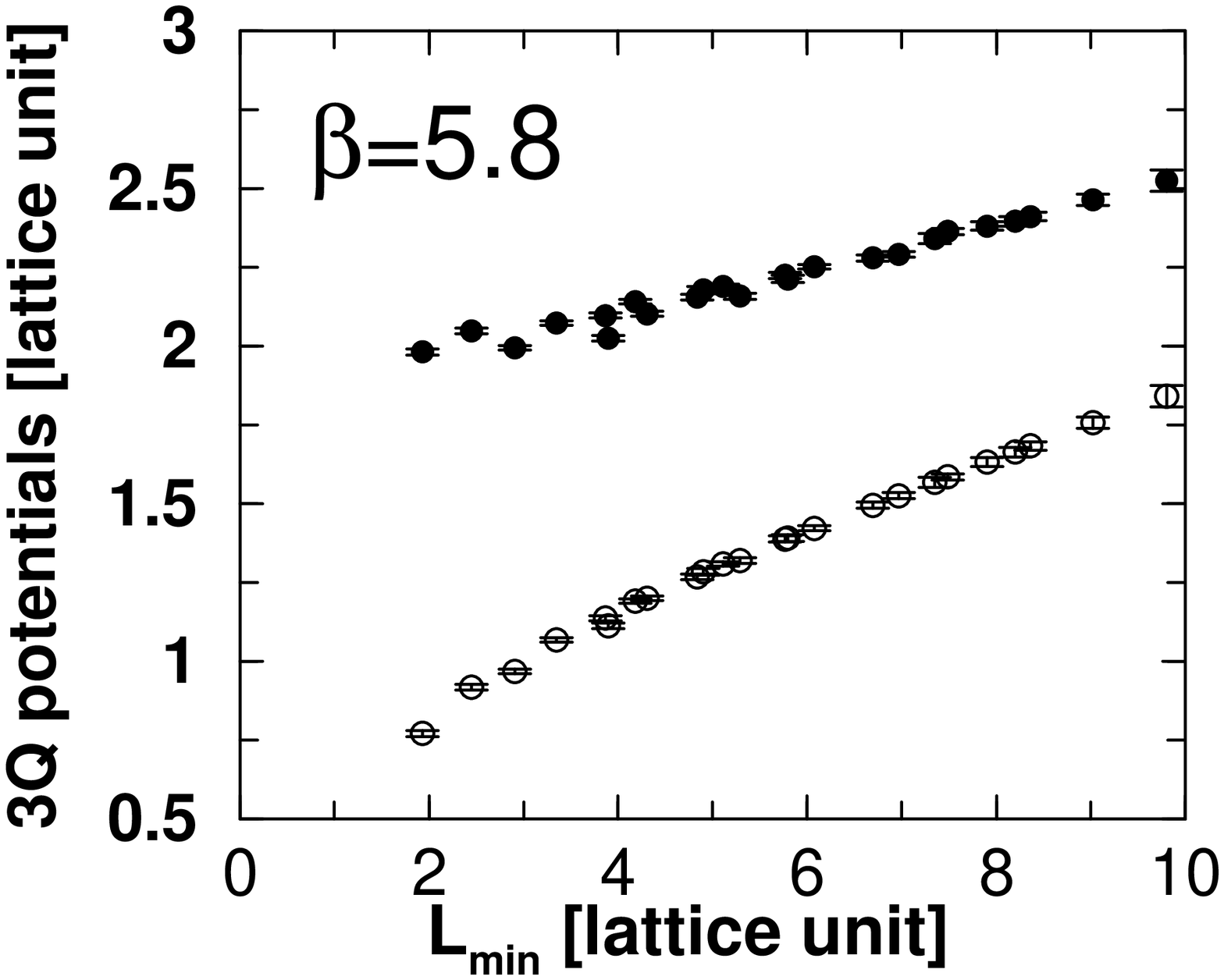}
\includegraphics[height=3.3cm]{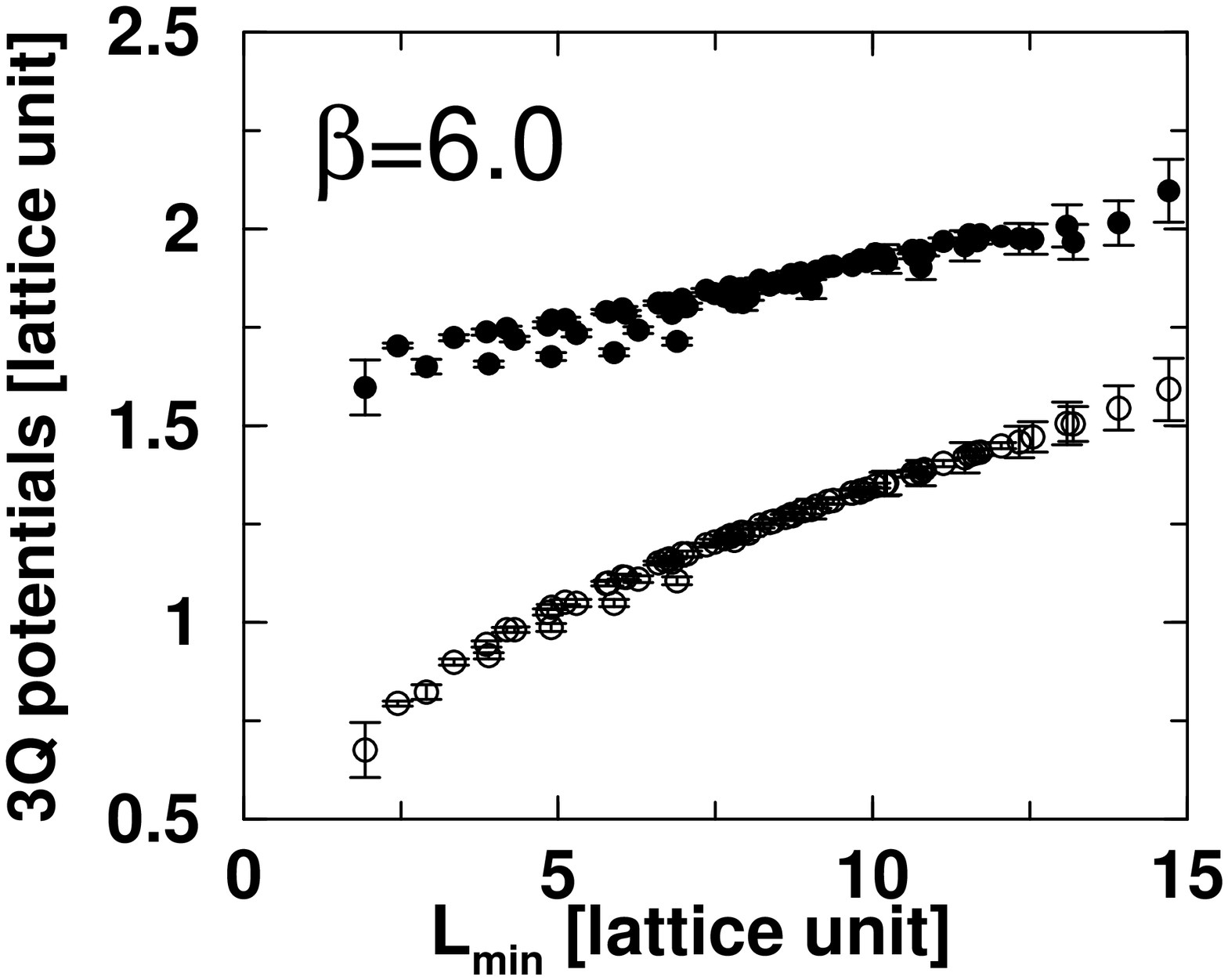}
}
\vspace{-0.8cm}
\caption{
The 1st excited-state 3Q potential $V_{\rm 3Q}^{\rm e.s.}$ and 
the ground-state 3Q potential $V_{\rm 3Q}^{\rm g.s.}$.
The lattice results at $\beta=5.8$ and $\beta=6.0$ well coincide 
besides an irrelevant overall constant. 
The gluonic excitation energy 
$\Delta E_{\rm 3Q} \equiv V_{\rm 3Q}^{\rm e.s.}-V_{\rm 3Q}^{\rm g.s.}$ 
is found to be about 1GeV in the hadronic scale.
}
\label{fig6}
\vspace{-0.4cm}
\end{figure}

The energy gap between $V_{\rm 3Q}^{\rm g.s.}$ and $V_{\rm 3Q}^{\rm e.s.}$ physically means 
the excitation energy of the gluon-field configuration in the presence of the spatially-fixed three quarks, 
and the gluonic excitation energy $\Delta E_{\rm 3Q} \equiv V_{\rm 3Q}^{\rm e.s.}-V_{\rm 3Q}^{\rm g.s.}$ 
is found to be about 1GeV \cite{TS03,STI04}
in the hadronic scale as 
$L_{\rm min}\sim 1~{\rm fm}$.

Note that the gluonic excitation energy of about 1GeV is rather large in comparison with 
the excitation energies of the quark origin.
The present result predicts that the lowest hybrid baryon, 
which is described as 
$qqqG$ in the valence picture, has a large mass of about 2 GeV \cite{TS03,STI04}. 
(The present result seems to suggest 
the constituent gluon mass of about 1GeV.)

\section{Behind the Success of the Quark Model}

Finally, we consider the connection between QCD and the quark model 
in terms of the gluonic excitation \cite{TS03,STI04}. 
While QCD is described with quarks and gluons, 
the simple quark model successfully describes low-lying hadrons 
even without explicit gluonic modes.
In fact, the gluonic excitation seems invisible in the low-lying hadron spectra, 
which is rather mysterious.

On this point, we find the gluonic-excitation energy to be about 1GeV or more, 
which is rather large compared with the excitation energies of the quark origin,
and therefore the effect of gluonic excitations is negligible 
and quark degrees of freedom plays the dominant role 
in low-lying hadrons with the excitation energy below 1GeV.

Thus, the large gluonic-excitation energy of about 1GeV gives the physical reason for 
the invisible gluonic excitation in low-lying hadrons, 
which would play the key role for the success of the quark model 
without gluonic modes \cite{TS03,STI04}. 

In Fig.7, we present a possible scenario from QCD to the massive quark model 
in terms of color confinement and dynamical chiral-symmetry breaking (DCSB) \cite{STI04}.

\begin{figure}[hb]
\vspace{-3cm}
\centerline{\includegraphics[width=4.2in]{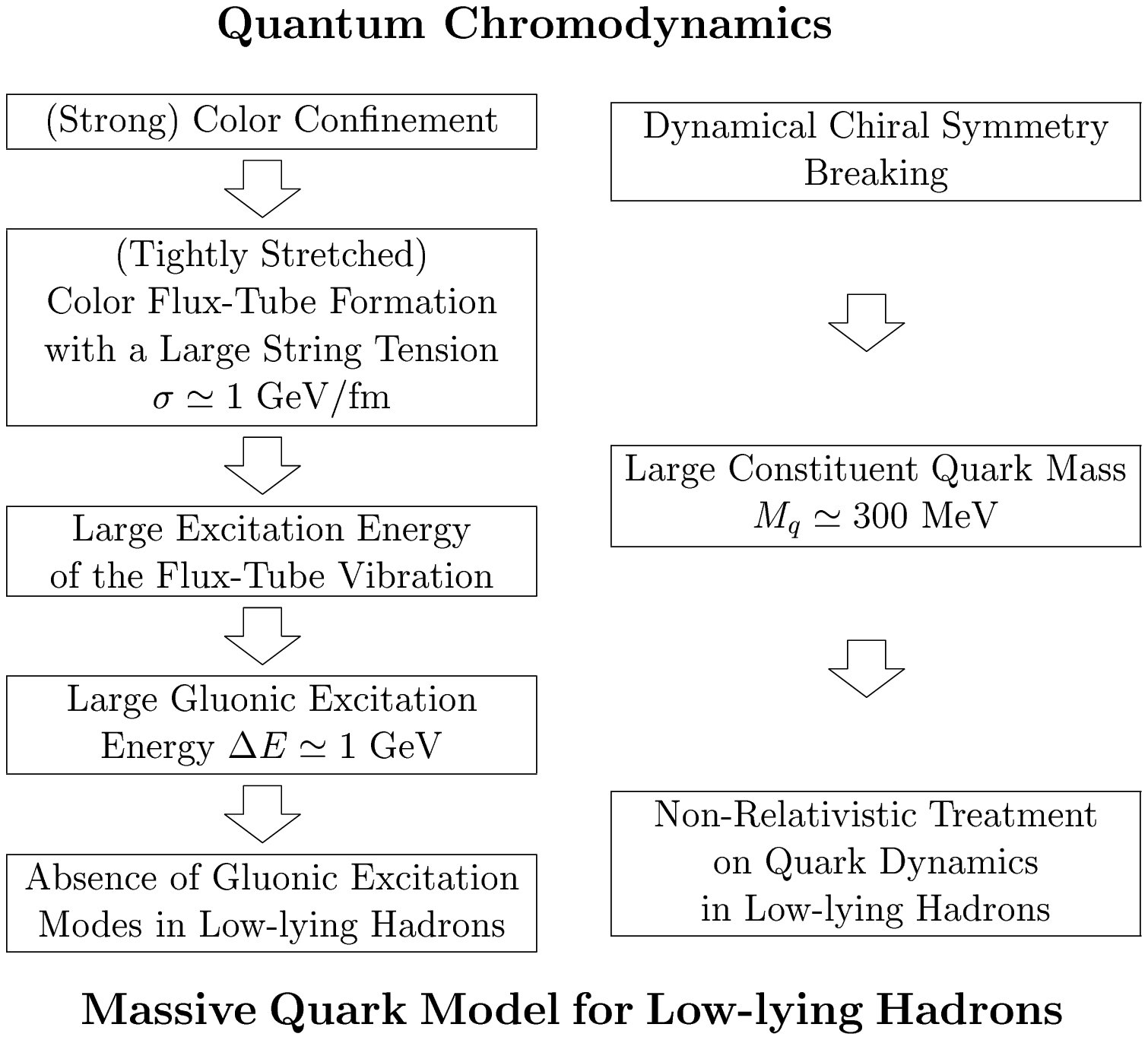}}
\vspace{-6.7cm}
\caption{A possible scenario from QCD to the quark model in terms of 
color confinement and DCSB.
DCSB provides a large constituent quark mass of about 300MeV, which enables the non-relativistic treatment for quark dynamics. 
Color confinement leads to the color flux-tube formation among quarks.
In the flux-tube picture, the gluonic excitation is described as the flux-tube vibration, 
and then its energy is expected to be large in the hadronic scale.
The large gluonic-excitation energy of about 1GeV leads to 
the absence of the gluonic mode in low-lying hadrons, 
which would play the key role to the success of the quark model without gluonic excitation modes.
}
\label{fig7}
\end{figure}


\begin{thebibliography}{9}
\bibitem{N66} Y.~Nambu, in {\it Preludes in Theoretical Physics}, 
(North-Holland, Amsteldam, 1966).
\bibitem{HN65} M.Y.~Han and Y.~Nambu, Phys. Rev. {\bf 139} (1965) B1006. 
\bibitem{conf2003} Articles in {\it Color Confinement and Hadrons in Quantum Chromodynamics}, 
edited by H.~Suganuma {\it et al.} (World Scientific, 2004).
\bibitem{NJL61} Y.~Nambu, G.~Juna-Lasinio, 
Phys. Rev. {\bf 122} (1961) 345; {\it ibid.} {\bf 124} (1961) 246.
\bibitem{R97} H.J.~Rothe, {\it Lattice Gauge Theories}, 
2nd edition (World Scientific, 1997) p.1.
\bibitem{TMNS01} T.T.~Takahashi, H.~Matsufuru, Y.~Nemoto,  H.~Suganuma, 
Phys. Rev. Lett. {\bf 86} (2001) 18.
\bibitem{TSNM02}T.T.~Takahashi, H.~Suganuma, Y.~Nemoto, H.~Matsufuru, 
Phys.Rev. {\bf D65} (2002) 114509.
\bibitem{TS03} T.T.~Takahashi and H.~Suganuma, Phys. Rev. Lett. {\bf 90} (2003) 182001.
\bibitem{TMNS99} T.T.~Takahashi, H.~Matsufuru, Y.~Nemoto and H.~Suganuma,   
{\it Dynamics of Gauge Fields}, Tokyo, Dec. 1999,  
edited by A.~Chodos {\it et al.}, (Universal Academy Press, 2000) 179; 
H.~Suganuma, Y.~Nemoto, H.~Matsufuru and T.T.~Takahashi, Nucl.Phys. {\bf A680} (2000) 159.
\bibitem{KS75CKP83} J.~Kogut, L.~Susskind, Phys. Rev. {\bf D11} (1975) 395;
J.~Carlson, J.~Kogut, V.~Pandharipande, Phys. Rev. {\bf D27} (1983) 233; {\bf D28} (1983) 2807.
\bibitem{AdFT02} C.~Alexandrou, P.~de~Forcrand, A.~Tsapalis, Phys. Rev. {\bf D65} (2002) 054503.
\bibitem{JdF04} O. Jahn and P. de Forcrand, Nucl. Phys. {\bf B} (Proc. Suppl.) {\bf 129} (2004) 700.
\bibitem{KS03} D.S. Kuzmenko and Yu.A. Simonov, Phys. Atom. Nucl. {\bf 66} (2003) 950.
\bibitem{C96} J.M.~Cornwall, Phys. Rev. {\bf D54} (1996) 6527.
\bibitem{C04} J.M.~Cornwall, Phys.Rev.{\bf D69} (2004) 065013.
\bibitem{IBSS03} H.~Ichie, V.~Bornyakov, T.~Streuer and G.~Schierholz, 
Nucl. Phys. {\bf A721} (2003) 899; Nucl. Phys. {\bf B} (Proc.Suppl.) {\bf 119} (2003) 751. 
\bibitem{STI04} H.~Suganuma, T.T.~Takahashi and H.~Ichie, 
{\it Color Confinement and Hadrons in Quantum Chromodynamics}, 
(World Scientific, 2004) p.249; Nucl.~Phys. {\bf A} (2004).
\bibitem{N6970} Y.~Nambu, in {\it Symmetries and Quark Models} (Wayne State University, 1969); 
{\it Lecture Notes at the Copenhagen Symposium} (1970).
\bibitem{H65} R.~Hagedorn, Nuovo Cim.Suppl. {\bf 3} (1965) 147.
\bibitem{N74} Y.~Nambu, Phys. Rev. {\bf D10} (1974) 4262.
\end{thebibliography}
\end{document}